\documentclass[amsmath,amssymb,showkeys]{revtex4-2}

\usepackage{amssymb}
\usepackage{amsmath}
\usepackage{amsthm}
\usepackage{csquotes}
\usepackage{graphicx}

\usepackage{pdfpages}
\makeatletter
\AtBeginDocument{\let\LS@rot\@undefined}
\makeatother

\usepackage{hyperref}
\hypersetup{
	colorlinks=true,
	linkcolor=blue,
	filecolor=blue,      
	urlcolor=blue,
	citecolor=blue
}

\usepackage{parskip}
\begin{document}
	
\title{Spectral taxonomy for quartic systems: fundamental clock, parity, and continuum}

\author{Teepanis Chachiyo}
\email{teepanisc@nu.ac.th}
\affiliation{Department of Physics, Faculty of Science, Naresuan University, Phitsanulok 65000, Thailand}

\begin{abstract}
A symmetric quartic potential is a physics model with expansive applications, ranging from broadband energy harvesters, quantum tunneling in molecules and the early universe, to torque-free spacecraft rotation. Its rich dynamics have been traditionally classified into regimes and expressed as disjointed time-domain solutions. Here we build a taxonomy for this broad class of motions and discover that their regimes exhibit a symmetrical spectral structure: they share a fundamental clock, obey parity selection, and dissolve into the separatrix through a discrete-to-continuum transition. Applied to the famous Dzhanibekov effect where a rotating body periodically undergoes rapid 180-degree flips in its attitude, the taxonomy reveals its spectral anatomy. The three principal-axis rotations share a common clock while occupying distinct parity channels, with stable-axis branches exchanging DC bias across the separatrix. We discuss a case study where the three spectral pillars: clock, parity, and continuum, survive the Wick rotation from real-time into imaginary-time kinematics.
\end{abstract}


\maketitle

\begin{center}
\includegraphics[width=\textwidth]{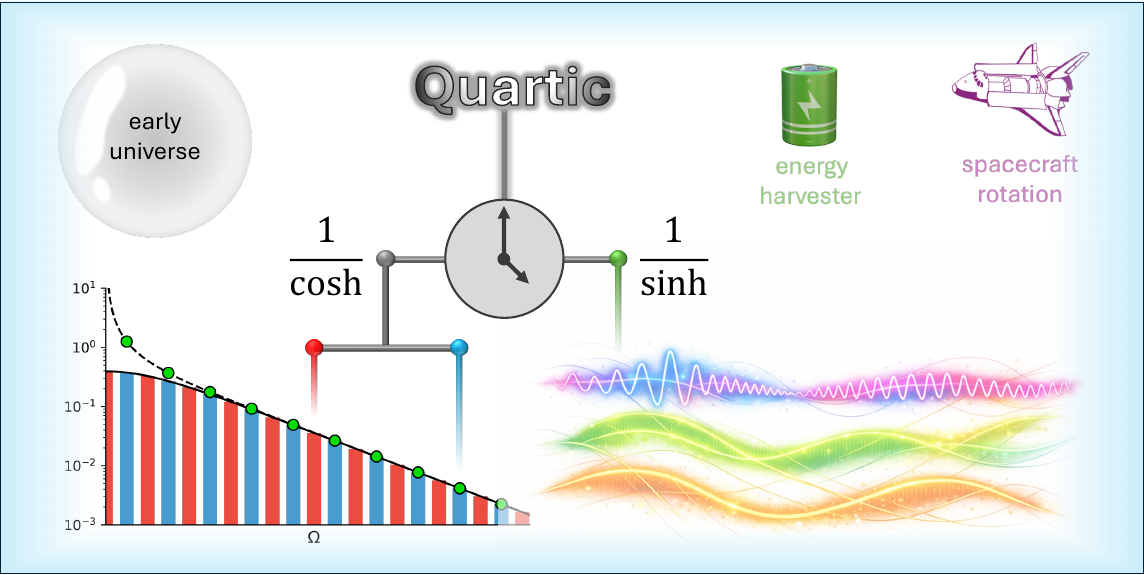}
Graphical Abstract
\end{center}

\section{Introduction}\label{sec:intro}

Analytical solutions for the quartic potential were introduced by Georg Duffing in 1918~\cite{duffing1918erzwungene}, using the machinery formalized nearly a century earlier by Carl Gustav Jacob Jacobi in his seminal 1829 treatise \emph{Fundamenta nova theoriae functionum ellipticarum}~\cite{jacobi1829fundamenta}. This centuries-long progression illustrates a broader perspective: physics is not merely about inventing new mathematics, but also deciphering a reflection of mathematics that reveals a hidden design of nature. Here, we stand on the shoulders of giants to complete this historical arc. We build a taxonomy, classify all possible solutions, and show that they have a symmetrical spectral structure---a unification that dictates the fundamental frequency and selection rules for both classical and quantum systems.

Recently, exact frequency-domain solutions across all regimes for pendulum-like systems have been derived via complex time~\cite{Chachiyo2025upl}. The study also finds a previously hidden symmetry: their dynamical regimes share a single spectral kernel, with parity selection distinguishing regimes, and the separatrix as the continuum limit. Here, we pursue this further and discover that the same symmetry persists in a much larger class of physics models---the quartic oscillator~\cite[\href{http://dlmf.nist.gov/22.19.ii}{$\S$22.19(ii)}]{NIST:DLMF}.

This paper is organized as follows. Section \ref{sec:intro} motivates the broader significance of the quartic potential. Section \ref{sec:taxonomy} introduces the taxonomy that reduces the infinite parameter space of the potential into only nine archetypes. Section \ref{sec:formalism} highlights the resulting solutions. Section \ref{sec:spacecraft} exemplifies the symmetrical spectral structure via the torque-free spacecraft rotation. Section \ref{sec:method} presents the mathematical derivations, and Section \ref{sec:discussion} discusses the Wick rotation into the imaginary time kinematics. Finally, Section \ref{sec:conclusion} summarizes our findings and proposes further research outlooks.

The symmetric quartic potential

\begin{equation}
U(x) = c_4 x^4 + c_2 x^2, \quad \left(\frac{dx}{dt}\right)^2 = 2 E - 2 U(x), \label{eq:quartic_physical}
\end{equation}
is a mathematical blueprint that transcends physical scales and disciplines.  In classical physics, the potential serves as the foundation for the Duffing oscillator~\cite{nayfeh2024nonlinear,kovacic2011duffing,SalasSalas2021,vonWagner2024}, which in turn dictates the mechanics of broadband vibration energy harvesting~\cite{Jia2020} and stochastic resonance~\cite{Gammaitoni1998}. In quantum physics, the quartic potential is a model to understand tunneling and quantum transport in molecules~\cite{Rastelli2012,Pathak2026m,Pathak2026u}. In cosmology, tunneling through a quartic potential represents the phase transition mechanism in the early universe~\cite{Adams1993}. Here, we briefly review these examples to demonstrate the incredibly expansive applications of the quartic potential.

In energy harvesting, the symmetric quartic potential has revolutionized the design of Vibration Energy Harvesters (VEHs)~\cite{Jia2020}. Traditional linear VEHs operate efficiently only when the ambient environmental vibration perfectly matches their single resonant frequency. If the ambient frequency shifts, the power output drops significantly. To overcome this, engineers construct nonlinear harvesters using magnetic repulsion, piezoelectric cantilevers, or tuned mass damper-inerters that physically emulate a symmetric bistable quartic potential~\cite{Pertin2022}. Because the effective stiffness of the quartic potential changes with the amplitude of displacement, the resonance peak of the harvester bends, creating a large operational bandwidth~\cite{Ferrari2009,Zou2021,Chen2021}.

In stochastic resonance, the symmetric quartic potential is configured as a bistable double-well ($c_2 < 0$)~\cite{Gammaitoni1998}. Stochastic Resonance (SR) is a counter-intuitive phenomenon wherein the addition of random noise actually \emph{enhances} the detection of a weak, periodic input signal~\cite{Zhang2015}.   The dynamics of SR are governed by the underdamped asymmetric periodic potential stochastic resonance (UAPPSR) model. The system is modeled by a Langevin equation incorporating the quartic potential, a weak periodic input, and Gaussian white noise~\cite{Hu2021}. If the input signal is too weak to push a particle over the central barrier of the quartic potential, the particle remains trapped in one well, rendering the signal undetectable. 

However, when optimal ambient noise is introduced, it provides the requisite kinetic energy for the particle to hop over the barrier. Because the weak periodic signal rhythmically modulates the relative depths of the two wells, the noise-induced hopping synchronizes with the signal frequency. This synchronization transfers energy from the broadband noise into the spectral peak of the signal, amplifying the Signal-to-Noise Ratio (SNR). This mechanism is actively deployed in industrial engineering for mechanical fault diagnosis, allowing sensors to detect the faint, low-frequency vibrations of a failing rotary bearing amidst the deafening background noise of factory machinery~\cite{Hu2021}.

In quantum physics, tunneling is one of the most profound departures from classical theory. In molecules, its study dates back to as early as 1927 and is commonly described using a double-well potential~\cite{Rastelli2012,Hund1927,TunnellingInMolecules2020}. In this framework, the wavefunction leaks from the wells into the barrier, causing the lowest energy levels to split into a closely spaced pair: a lower-energy symmetric ground state and a higher-energy antisymmetric excited state. The energy difference, defined as the tunneling splitting ($\Delta E$), is inversely related to the tunneling time~\cite{Ferro-Costas2020}. Historically, this model was essential for explaining the inversion spectrum of ammonia, but its scope has since broadened to include phenomena such as hydrogen bonding, proton transfer, and complex polyatomic rearrangements~\cite{Rastelli2012,Pathak2026m,Pathak2026u,Erakovi2022}.

In cosmology, quantum tunneling through a quartic potential is the primary engine for first-order phase transitions in the early universe, governing processes such as early inflation and electroweak symmetry breaking~\cite{Coleman1977,Coleman1980,Adams1993}. As the primordial universe cooled, its underlying scalar fields often became temporarily trapped in an elevated, metastable energy state known as a false vacuum. To achieve true stability, these fields had to quantum-tunnel through an energy barrier to the absolute minimum, or true vacuum. In particular, detailed calculations of tunneling rates for scalar fields with quartic potentials have been developed~\cite{Adams1993}. This tunneling process creates a small region in space where the field sits in the true vacuum, surrounded by false vacuum---a bubble of true vacuum.

Once formed, sufficiently large bubbles expand rapidly, filling the early universe with an ensemble of growing bubbles as the phase transition proceeds. The precise rate at which these bubbles form and collide is critical: it determines whether the universe can successfully complete its transition before expanding too far. Consequently, calculating this tunneling rate accurately is essential for resolving the classic \textquote{graceful exit} problem~\cite{Guth1981,Guth1983}, providing the mechanism by which the early universe could smoothly terminate its rapid expansion and settle into the hot, matter-filled state we observe today.

In astrodynamics (e.g., a satellite adrift in space), the differential equations governing the torque-free rigid-body 3D rotation exhibit an exact mathematical homology to the symmetric quartic potential~\cite{Landau1976,Goldstein1980,Arnold1989}. The torque-free motion of an arbitrary rigid body is governed by Euler's equations of motion in which two fundamental quantities are conserved: rotational kinetic energy ($K = \frac{1}{2} I_1 \omega_1^2 + \frac{1}{2} I_2 \omega_2^2 + \frac{1}{2} I_3 \omega_3^2$) and the square of the angular momentum ($L^2 = I_1^2 \omega_1^2 + I_2^2 \omega_2^2 + I_3^2 \omega_3^2$). These two conserved quantities effectively reduce the degrees of freedom for each of the rotation axes to only 1D dynamics, precisely the quartic form (i.e., $U(\omega_1) = c_4 \omega_1^4 + c_2 \omega_1^2$).

These quartic dynamics govern the stability of the rigid body's rotation---the famous Dzhanibekov effect (or the Tennis Racket Theorem), where the rotating body periodically undergoes rapid 180-degree flips in its attitude~\cite{Murakami2016,Deriglazov2024,Peterson2021}.

The Dzhanibekov effect is actively being studied for applications in future space missions~\cite{TRIVAILO2019}. For example, a 6-mass model can be designed to switch ON/OFF the spacecraft flipping motion. A \textquote{Geometric Morphing} of spacecraft can be achieved by actively extending or contracting the arm lengths of these masses, thus stabilizing the spacecraft rotation.

In summary, the quartic potential is a mathematical blueprint that transcends physical scales and disciplines. By presenting the exact spectral solutions and their taxonomy, this paper offers a new frequency-aware framework by which physical systems can be characterized, designed, and controlled.

\section{Spectral Taxonomy}\label{sec:taxonomy}

To build a taxonomy, one must collapse the infinite parameter space of $(c_4, c_2)$ into a discrete set of scenarios. We propose using a linear transformation (see Appendix \ref{sec:scaling})

\begin{equation}
	q =  \xi x, \quad \tau = \lambda t, \quad \epsilon = \nu E. \label{eq:unit_scaling}
\end{equation}
to map the physical space into an entirely dimensionless topological space---the quartic backbone

\begin{equation}
	V(q) = \frac{1}{2} \sigma q^4  + \gamma q^2, \quad \left(\frac{dq}{d\tau}\right)^2 = 2 \epsilon - 2 V(q), \quad \sigma, \gamma \in \{-1, 0, +1\}. \label{eq:quartic_backbone}
\end{equation}

Via this strategic choice of spatial, temporal, and energetic units, all possible solutions collapse to nine dimensionless archetypes: $\sigma, \gamma \in \{ -1, 0, +1\}$.

\begin{figure}[h]
\begin{center}
	\includegraphics[width=\textwidth]{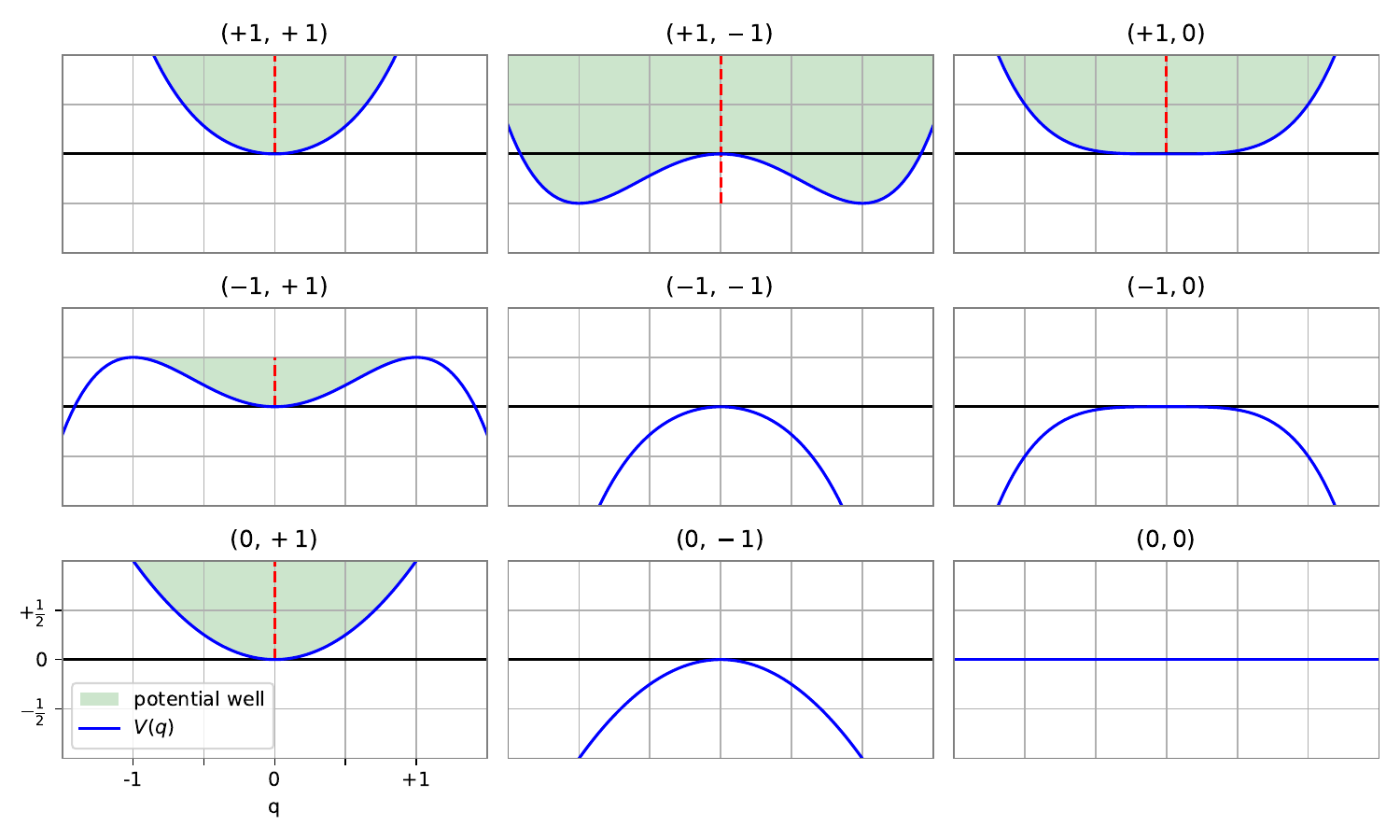}
\end{center}
	\caption{\textbf{Nine irreducible quartic archetypes $\sigma,\gamma \in \{-1,0,+1\}$:}  This includes the four bound, nontrivial archetypes which parallel the signum of $(c_4, c_2)$, one trivial simple harmonic motion, and four other unbound cases. The shaded areas indicate the bound motions, while the dash-lines represent the range of energy $\epsilon$ in which the motion remains oscillatory.} \label{fig:potential}
\end{figure}

Figure~\ref{fig:potential} shows the complete taxonomy of the quartic backbone. The four lower-right archetype group does not exhibit oscillatory behavior, rather a runaway motion $q \to \pm \infty$.

Only two cases contain a separatrix: $(+,-)$ and $(-,+)$. For $(+,-)$, the separatrix is when the energy is just enough to cross the middle barrier $(\epsilon = 0)$, and the system transits from a double well to a global single well. For $(-,+)$, the separatrix is when the energy is just enough to cross the outer barrier $(\epsilon = +1/2)$, and the system transits from the middle pocket into a runaway motion.

We now focus on the four bound, nontrivial archetypes (the shaded panels in Fig.~\ref{fig:potential}, excluding the simple harmonic oscillation where $\sigma,\gamma = 0,+1$), exposing their regimes and separatrices.

As the quartic backbone follows a trajectory $q(\tau)$---oscillating forward and back---the turning points are the locations at which the velocity $\dot{q} = 0$. Let us define the characteristic squared-amplitudes $A_1, A_2$ found simply by factoring a polynomial. From Eq.~(\ref{eq:quartic_backbone})

\begin{equation}
\left(\frac{dq}{d \tau}\right)^2 = A_0 (q^2-A_1) (q^2 - A_2), 
\end{equation}
where 

\begin{equation}
A_0 = -\sigma, \quad  A_{1,2} = \frac{ -\gamma \pm  \sqrt{\gamma^2 + 2 \sigma \epsilon} }{\sigma}.\label{eq:roots_backbone}
\end{equation}

The turning points are located at $\pm\sqrt{A_{1,2}}$ when $A_{1,2} > 0$. These points are typically called roots and ultimately depend on the energy $\epsilon$ (or equivalently the initial condition) of the motion. Here, we classify all oscillatory and separatrix motions of the quartic backbone, providing the exact trajectory $q(\tau)$ via a spectral lens.

\section{Formalism}\label{sec:formalism}

For a given bound, nontrivial $(\sigma, \gamma)$, the constants  $A_{1,2}(\epsilon)$ can be evaluated at a fixed energy. The spectral decomposition then provides the trajectory $q(\tau)$, whose derivatives promptly yield the velocity, acceleration, and other related quantities of interest. The taxonomy of solutions $q(\tau)$ consists of two classes, distinguished solely by $\sigma$:

\begin{equation}
\begin{aligned}
& \text{Class 1, } \sigma = +1: \\
& \qquad k < 1, \; q(\tau) =  \sum_{n \text{ odd}}^\infty c_n \cos(n \Omega_0 \tau), \\
& \qquad k > 1, \; q(\tau) =  \sum_{n \text{ even}}^\infty c_n \cos(n \Omega_0 \tau) + \frac{c_0}{2}, \\
&\text{Class 2, } \sigma = -1: \\
& \qquad k < 1, \; q(\tau) =  \sum_{n \text{ odd}}^\infty s_n \sin(n \Omega_0 \tau).
\end{aligned}\label{eq:spectral_osc}
\end{equation}

Here, the index $n=0$ is always omitted from the summation. The separatrices, sitting at the boundary $k \to 1$, are the continuum limit of the oscillatory regimes:

\begin{equation}
	\begin{aligned}
		\text{Class 1:} &  \;\; k \to 1, \;  q(\tau) = \int_0^\infty d\Omega \, C(\Omega) \cos(\Omega \tau)  \\
		\text{Class 2:} & \;\; k \to 1, \;  q(\tau) = \int_0^\infty d\Omega  \, S(\Omega) \sin(\Omega \tau)  
	\end{aligned}\label{eq:spectral_sep}
\end{equation}

The spectral coefficients and densities are
\begin{equation}
	\begin{aligned}
		c_n & = \frac{2  \Omega_0}{\zeta \cosh(\kappa \frac{n \Omega_0}{\Omega_L})}, \quad    & s_n &= \frac{2  \Omega_0}{\zeta \sinh(\kappa \frac{n \Omega_0}{\Omega_L})}, \\
		C(\Omega) & = \frac{1}{\zeta \cosh(\kappa \frac{\Omega}{\Omega_L})}, \quad  & S(\Omega) & = \frac{1}{ \zeta \sinh(\kappa \frac{\Omega}{\Omega_L})}
	\end{aligned}
\end{equation}

Taken together, the taxonomy shows all possible bound motions of the quartic backbone via a spectral lens. There are a number of dynamical spectral parameters. These are space-time scale $\zeta$, frequency scale $\Omega_L$, elliptic modulus $k$, period $T$, fundamental frequency $\Omega_0$, and envelope scale $\kappa$. Once the square-amplitudes $A_{1,2}(\epsilon)$ have been evaluated for a fixed energy $\epsilon$, these parameters can be evaluated in succession as follows:

\begin{equation}
\begin{aligned}
 \text{space-time scale } & \zeta && = \sqrt{|A_0|} \\
 \text{frequency scale } & \Omega_L && = \begin{cases} \text{Class 1: } \quad   \zeta \sqrt{A_1-A_2} \\
 \text{Class 2: } \quad \zeta \sqrt{A_2} \end{cases}
\end{aligned} \label{eq:spectral_scale}
\end{equation}

The frequency scale $\Omega_L$ is defined specifically for each class, while the other dynamical parameters are defined uniformly across all regimes and all classes.

\begin{equation}
\begin{aligned}
	\text{elliptic modulus } & k && =  \zeta {\sqrt{A_1}}/{\Omega_L}, \\
	\text{fundamental period } & T && =  {4 \Re[K(k)]}/{\Omega_L} \\
	\text{fundamental frequency } &   \Omega_0 && = 2\pi / T, \\
	\text{envelope scale } & \kappa && = K'(k), \\
\end{aligned} \label{eq:spectral_params}
\end{equation}

For simplicity, the solutions given above are consistent with a suitable choice of initial position and velocity. Specifically, $\{q_0,\dot{q}_0\} = \{   \sqrt{A_1},0\}$ for Class 1, and $\{0,\sqrt{2 \epsilon}\}$ for Class 2. To accommodate other choices of initial conditions, the time can be shifted $q(\tau) \to q(\tau-\tau_0)$ with an additional overall sign $q \to \pm q$ applied when needed.

The only special function needed for this spectral framework is the complete elliptic integral of the first kind $K(k)$, which can be evaluated by the SciPy library~\cite{2020SciPy-NMeth} in Python. Alternatively, an efficient and accurate approximation for the full range of modulus, $K \approx \frac{1}{n}\ln[(\frac{4}{\sqrt{1-k^2}})^n+b]$, is suitable  for platforms that special functions are not accessible such as web-based and firmware developments, with constants $n$ and $b$ defined in Ref.~\cite{Chachiyo2026} (see supplementary material for implementation details).

\section{Spacecraft Rotation: Clock, Parity, and Continuum}\label{sec:spacecraft}

The Dzhanibekov effect is a celebrated phenomenon in which a torque-free rigid body---such as a tumbling spacecraft or a flipped tennis racket---undergoes a bizarre 180-degree flip after a lingering interval of stable rotation. Although the precise time-domain solutions for all three rotational axes are well-known~\cite{Landau1976}, the present formalism offers a previously hidden perspective. For any torque-free rigid body---regardless of its shape or size---all principal axes simultaneously share a fundamental clock, obey parity selection, and dissolve into the continuum limit at the separatrices.

\begin{figure}[h]
	\includegraphics[width=\textwidth]{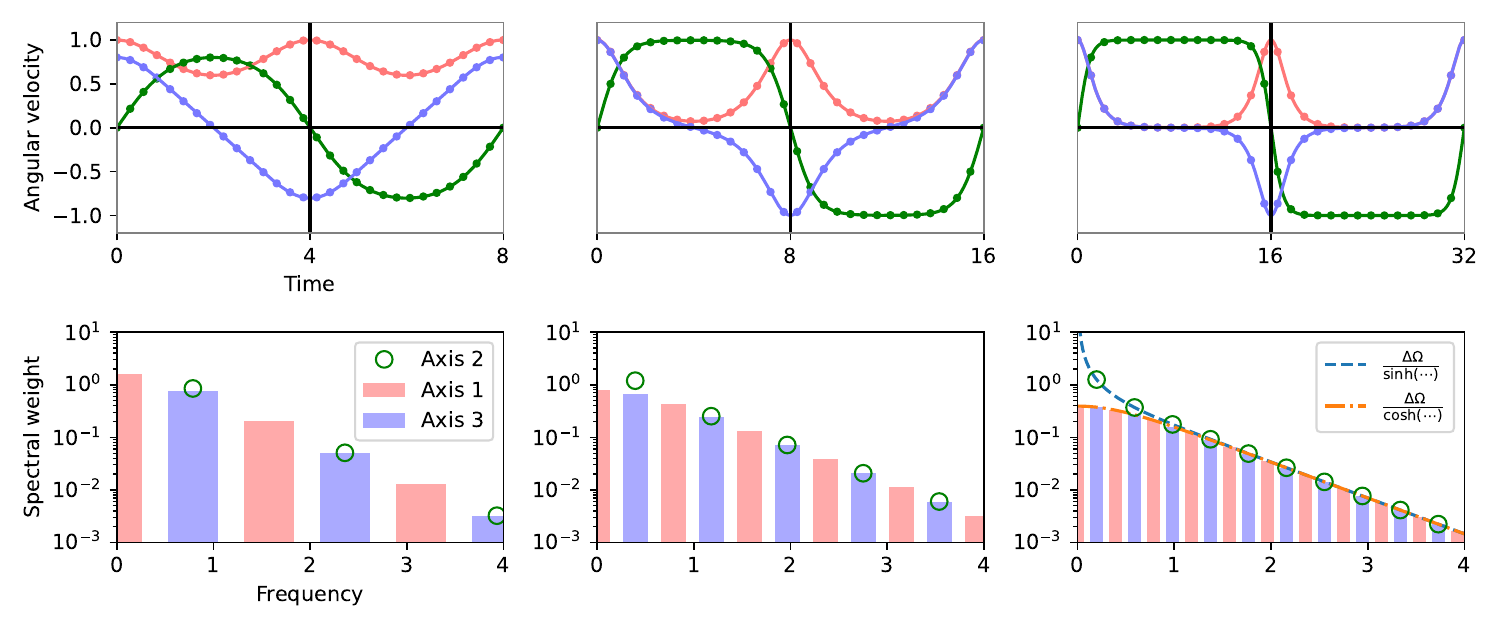}
	\caption{\textbf{Rotation around the principal axes of a spacecraft:} (Upper panels) showing normalized angular velocity from numerical solutions (solid-line), matching the present analytical spectral solutions (dot) for three sets of normalized energy and angular momentum $C_1 = 1, C_3 = k^2$, where the modulus $k < 1$ is chosen such that the periods are $8$, $16$, and $32$ seconds, respectively. (Lower panels) showing their spectral decomposition. The color representations are (red, green, blue) for the rotation about (major, intermediate, minor) axis, respectively. }   \label{fig:spacecraft}
\end{figure}

The spectral taxonomy (see Method \ref{sec:method_spacecraft}) gives the fundamental clock:

\begin{equation}
T = \frac{4 K(\sqrt{C_\text{min}/C_\text{max}})}{\sqrt{C_\text{max}}}. \label{eq:Dzhanibekov_period}
\end{equation}

Here, $C_\text{min}, C_\text{max}$ are the smaller and larger of the two constants,

\begin{equation}
	\begin{aligned}
C_1 & = \left( \frac{I_1 - I_2}{I_1 I_2 I_3} \right) (L^2 - 2 E I_3), \\
C_3 & = \left( \frac{I_2 - I_3}{I_1 I_2 I_3} \right) (2 E I_1 - L^2).
	\end{aligned} \label{eq:C1C3}
\end{equation}

Its continuum limit simplifies to the well-known \textquote{Dzhanibekov condition}.

\begin{equation}
C_1 = C_3 \quad \Longleftrightarrow \quad E = \frac{L^2}{2 I_2}.
\end{equation}

Figure~\ref{fig:spacecraft} introduces this clock visually. In the upper panels, the normalized angular velocities of the three principal axes are shown as functions of time. As a validation, the numerical trajectories are overlaid with the spectral taxonomy solutions. In the lower panels, the same motions are decomposed into their frequency components. Moving from left to right, the system approaches the separatrix, the period lengthens, and the discrete spectral lines become more closely spaced. Thus, before discussing the individual consequences, Fig.~\ref{fig:spacecraft} already shows the main spectral picture: all three principal-axis components share one clock, occupy distinct parity channels, and approach a continuum spectrum near the Dzhanibekov threshold.

In the traditional description, the Dzhanibekov effect is identified through the intermediate-axis separatrix: the condition tells us where the flip occurs. The spectral formulation adds a complementary answer: it tells us how the motion is organized, as summarized by the following consequences.

\subsection{The traditional Dzhanibekov condition locates the flip; the present formulation reveals its spectral anatomy.}

The classical criterion \(E=L^2/(2I_2)\) identifies the separatrix of the intermediate-axis instability. The spectral taxonomy goes further: it shows how the three principal axis angular velocity components reorganize around this boundary through a shared clock, parity-selected harmonics, and a discrete-to-continuum transition. In this view, the Dzhanibekov effect is not only a geometric instability of the intermediate axis, but a global spectral rearrangement of torque-free rigid-body motion.

Revisit closely the overall feature of Fig. \ref{fig:spacecraft}. In the upper panels, the three lines---red, green, blue---are perfectly synchronized. Going from left to right, even after varying angular momentum and energy to adjust the period of rotation, every principal-axis strictly and simultaneously shares a fundamental clock.

In the lower panels, their spectral anatomies are obvious. Following the same color representation---red, green, blue---each axis organizes itself into a distinct parity occupation. Going from left to right, even after varying angular momentum and energy, the parity remains a unique character of the torque-free rigid body rotation.

Taxonomic analysis in Eq.~(\ref{eq:proof_space_class}) and Eq.~(\ref{eq:proof_space_modulus}) shows that the rotation of Axis 1 (major) and Axis 3 (minor) are exclusively of Class 1.  The relative strength between the normalized energy and angular momentum (i.e., $C_1 < C_3$ and vice versa) determines which axis occupies odd or even parity in the frequency space. Nevertheless, the parity split is always observed at all times, regardless of shape or size of the rigid body.

In the lower panels, the rightmost one represents the limit in which the separatrix is about to occur. The red--green--blue dots and bars are squeezed tightly packed together, forming continuous samplings in the frequency space. The Fourier coefficients of the oscillatory regimes are matching perfectly the Fourier transform of the separatrices themselves. In addition to the pendulum-like dynamics~\cite{Chachiyo2025upl}, the quartic systems contain two kernels with strikingly parallel forms:

\begin{equation}
	\text{Class 1: } \; \frac{1}{\cosh(\kappa \frac{\Omega}{\Omega_L})}, \qquad \text{Class 2: } \; \frac{1}{\sinh(\kappa \frac{\Omega}{\Omega_L})}.
\end{equation}

Taken together, Fig. \ref{fig:spacecraft} highlights the symmetrical spectral structure in quartic systems where regimes share a fundamental clock, obey parity selection, and merge into a continuous spectrum when the oscillating dynamics transit into their stopping motions (separatrix).

\subsection{The intermediate component has no DC bias}

For the intermediate axis \(y_2\), the governing quartic is always Class 2:
\begin{equation}
	y_2(t)=\sum_{n \text{ odd}} s_n \sin(n\Omega_0 t),
\end{equation}
and therefore
\begin{equation}
	\langle y_2\rangle =0.
\end{equation}

Thus, the intermediate axis is distinguished by the absence of a DC component. Its motion is carried entirely by odd sine harmonics, so over a full clock cycle, the angular velocity must change sign rather than remain biased toward one direction.

This structure is visible in Fig.~\ref{fig:spacecraft} (upper-right corner), where the system is close to the separatrix and the tennis-racket effect is active. During the lingering interval, the green \(y_2\) component remains on its highest plateau while the other two are near zero. In physical space, this corresponds to the angular velocity vector \(\vec{\omega}\) pointing predominantly along the intermediate principal axis.

The lingering occurs because the system is approaching the peak of its potential energy barrier, while being under a near-separatrix condition. This stopping motion is best understood via the pendulum-like analogy~\cite{Chachiyo2025upl}. Near the separatrix, a pendulum only has enough total energy to reach the top. However, as the mass approaches the top, it can never quite get there. This is because its kinetic energy is spent climbing the potential peak, thus slowing it down asymptotically to a halt. In pendulum-like dynamics, this lingering is called the Stopping motion. Once the trajectory leaves this potential peak, the component $y_2$ decays rapidly and changes sign, producing the apparent sudden flip of the body.

In this view, the flip is not an anomalous interruption of rotation. It is the visible consequence of a zero-bias spectral branch: the intermediate component cannot keep a permanent sign, and near the separatrix this sign reversal is stretched into a long pause followed by a rapid transition.

\subsection{The major and minor components form complementary Class-1 regimes}

As shown in the classification analysis (see Method  \ref{sec:method_spacecraft}), the major- and minor-axis components split into two complementary scenarios, depending on the relative magnitude of the normalized angular momentum $C_1$ and energy $C_3$:
\begin{equation}
	\begin{aligned}
		& C_1 > C_3: \begin{cases}
			y_1 \text{: Class 1}, \; k > 1 \quad & \text{even} \text{ harmonics + DC}, \\
			y_3 \text{: Class 1}, \; k < 1 \quad & \text{odd } \text{ harmonics},
		\end{cases} \\
		& C_3 > C_1: \begin{cases}
			y_1 \text{: Class 1}, \; k < 1 \quad & \text{odd } \text{ harmonics}, \\
			y_3 \text{: Class 1}, \; k > 1 \quad & \text{even} \text{ harmonics + DC}.
		\end{cases}
	\end{aligned}
\end{equation}

Thus, the two stable-axis components are spectral counterparts: one carries the DC-biased even branch, while the other carries the zero-mean odd branch. Crossing the separatrix at $C_1=C_3$ swaps these assignments, transferring the DC-biased spectral role from one stable axis to the other. In this sense, the separatrix does not merely divide two types of motion; it exchanges the spectral identities of the major and minor components.

\subsection{A DC component signals persistent directional bias}

For the Class-1 \(k>1\) branch,
\begin{equation}
	y_1(t)=\sum_{n \text{ even}} c_n \cos(n\Omega_0 t)+\frac{c_0}{2},
\end{equation}
so
\begin{equation}
	\langle y_1(t)\rangle=\frac{c_0}{2}\ne0.
\end{equation}

Thus, the DC term is the spectral marker of persistent rotation about a stable principal direction. Unlike a zero-mean component, which alternates symmetrically between signs, a DC-biased component remains centered on one side of zero. In physical space, this bias keeps the angular velocity vector \(\vec{\omega}\) predominantly aligned with the corresponding principal axis.

For example, Fig.~\ref{fig:spacecraft} (upper-left corner) shows the \(y_1\) component fluctuating around a positive mean value, giving the visual impression of sustained rotation about the \(y_1\) axis.

\subsection{Parity selection gives engineering leverage} 

The parity rules turn the spectral taxonomy into an engineering design tool. Because each angular velocity component occupies only selected harmonics of the shared clock, disturbances do not couple to all frequencies equally. Structural modes, actuator inputs, and flexible appendage dynamics interact most strongly with the allowed parity channels, while forbidden harmonics are naturally suppressed. This converts torque-free tumbling from a purely time-domain crisis into a frequency-domain design opportunity.

The spectral taxonomy provides a harmonic map for resonance avoidance. Dynamic perturbations (e.g., actuator inputs and flexible appendages) tuned near allowed parity channels may couple strongly to the tumbling motion, whereas modes tuned to absent harmonics are naturally suppressed in the ideal symmetric model. In this sense, the Dzhanibekov motion becomes a frequency-design problem: one can identify which harmonics should be avoided, targeted, or shifted by changing the spacecraft geometry, inertia distribution, or control strategy.

If a spacecraft changes its inertia distribution, for example by extending appendages or moving internal masses, then the principal moments \(I_1,I_2,I_3\) change. Consequently, the invariants \(C_1\) and \(C_3\), and hence the spectral parameters, $\Omega_0,\; k,\; \kappa$ are shifted. Thus, dynamically modulating the inertia distribution moves the spacecraft through the spectral taxonomy. Extending or retracting appendages does not merely change the rotational period; it can alter the harmonic spacing, transfer the DC-biased branch from one stable axis to the other, suppress or expose selected parity channels, and drive the system toward or away from the continuum limit.

\section{Method}\label{sec:method}

\subsection{Spectral Taxonomy and Symmetrical Spectral Structure}

Here, we derive the spectral taxonomy. The derivation initially follows the Jacobi elliptic framework where the Fourier series is expressed in terms of the elliptic nome $q(k) = e^{-\pi K'(k)/K(k)}$. However, we then need to break away from the nome formalism to separate out the characteristic frequency $\Omega_0$, which was previously entangled inside the exponent of the nome. Only then can the fundamental clock, parity selection, and continuum limit begin to emerge.

\subsubsection{Emergence of Frequency Scale and Modulus}

For the parameters $\Omega_L$ and $k$ to appear naturally, we start from the quartic backbone motion:

\begin{equation}
	\left( \frac{d q}{d \tau} \right)^2 = A_0 (q^2-A_1)(q^2-A_2),
\end{equation}
which can be written in an integral form:

\begin{equation}
	d \tau  = \frac{\pm dq}{\sqrt{A_0 (q^2-A_1)(q^2-A_2)}}.
\end{equation}

By change of variable $q \to \phi$, it is possible to map the above expression into the standard definition of the elliptic integral:

\begin{equation}
	d \tau = \frac{1}{\Omega_L}\frac{d\phi}{\sqrt{1-k^2\sin^2 \phi}},
\end{equation}
for which $\Omega_L$ and $k$ emerge as functions of $A_0, A_{1,2}$.

Because Class 1 and Class 2 have the opposite signs of $A_0$, they require different mapping:

\begin{equation}
	\begin{aligned}
		\text{Class 1:} & \quad q = \sqrt{A_1} \cos \phi \Rightarrow  d \tau = &  \frac{1}{\sqrt{|A_0|(A_1-A_2)}} & \frac{d\phi}{\sqrt{1- \frac{A_1}{A_1-A_2}\sin^2 \phi}}  \\ 
		\text{Class 2:} & \quad q =  \sqrt{A_1} \sin \phi \Rightarrow d \tau = &  \frac{1}{\sqrt{|A_0|A_2}}  & \frac{d\phi}{\sqrt{1-\frac{A_1}{A_2}\sin^2 \phi}}   
	\end{aligned}
\end{equation}

From these mappings, $\Omega_L$ can be identified as $\{\sqrt{|A_0|(A_1-A_2)},\sqrt{|A_0| A_2}\}$ and interpreted as the frequency under linear oscillation when $k \to 0$, hence the subscript $L$. Obtaining the explicit form of the frequency scale, we arrive at the elliptic modulus $k = \sqrt{|A_0| A_1}/\Omega_L$.

\subsubsection{The Jacobi Time-Domain Solutions}

With the explicit differential mappings established, integrating the equations of motion yields $\Omega_L \tau = F(\phi, k)$, where $F(\phi, k)$ is the incomplete elliptic integral of the first kind. Inverting this relationship defines the Jacobi amplitude function, $\phi = \text{am}(\Omega_L \tau, k)$. By substituting this amplitude back into our original phase mappings, the coordinates elegantly collapse into the standard Jacobi elliptic functions.

\begin{equation}
	\begin{aligned}
		q(\tau) & = \sqrt{A_1} \cos(\text{am}(\Omega_L \tau, k)) \equiv \sqrt{A_1} \text{cn}(\Omega_L \tau, k) \\
		q(\tau)  & = \sqrt{A_1} \sin(\text{am}(\Omega_L \tau, k)) \equiv \sqrt{A_1} \text{sn}(\Omega_L \tau, k)
	\end{aligned}
\end{equation}

For regimes in which $k > 1$, the transformative properties of the elliptic functions yield the exact solutions and their respective primitive periods as follows.

\begin{equation}
	\begin{aligned}
		\text{Class 1: } & \; k < 1,  \;\; q(\tau) = \sqrt{A_1}\text{cn}(\Omega_L \tau,k), && T_\text{prim} = 4 \Re[K(k)]/\Omega_L\\
		& \; k > 1, \;\; q(\tau) = \sqrt{A_1}\text{dn}(k \, \Omega_L \tau, \frac{1}{k}) && T_\text{prim} = 2 \Re[K(k)]/\Omega_L  \\ 
		\text{Class 2: } & \; k < 1, \;\; q(\tau) = \sqrt{A_1} \text{sn}( \Omega_L \tau,  k ) && T_\text{prim} = 4 \Re[K(k)]/\Omega_L
	\end{aligned}\label{eq:jacobi_sols}
\end{equation}

The above set of solutions exemplifies the fragmented nature of the standard Jacobi elliptic framework, treating oscillatory regimes and separatrix as fundamentally distinct entities governed by different mathematical rules and time scales. Here, we resolve this apparent division through a spectral lens.

\subsubsection{Fundamental Clock and Parity Selection}

Traditionally, the Fourier series for the Jacobi elliptic functions is presented in terms of the elliptic nome $q(k) = e^{-\pi K'(k)/K(k)}$. From the DLMF library~\cite[\href{http://dlmf.nist.gov/22.2}{$\S$22.2}]{NIST:DLMF}, the series can be written directly as follows:

\begin{equation}
	\begin{aligned}
		\text{Class 1: } & k < 1, \;\; q(\tau) = \sqrt{A_1} \frac{2\pi}{K(k) k} \sum_{n=0}^\infty \frac{q(k)^{n+1/2}}{1+q(k)^{2n+1}}\cos[(2n+1)\frac{\pi \Omega_L \tau}{2 K}] \\
		&  k > 1, \;\; q(\tau) =  \sqrt{A_1} \frac{2\pi}{K(\frac{1}{k})}\sum_{n=1}^\infty \frac{q(\frac{1}{k})^n}{1+q(\frac{1}{k})^{2n}}\cos[ 2 n \frac{\pi k \Omega_L \tau}{2 K(\frac{1}{k})}]  + \sqrt{A_1} \frac{\pi}{2 K(\frac{1}{k})}  \\ 
		\text{Class 2: }  & k < 1, \;\; q(\tau) = \sqrt{A_1} \frac{2 \pi}{K(k) k} \sum_{n = 0}^\infty \frac{q(k)^{n+1/2}}{1-q(k)^{2 n + 1}} \sin[ (2 n + 1) \frac{\pi \Omega_L \tau}{2 K(k)}]
	\end{aligned}
\end{equation}

However, for the fundamental clock, parity selection, and continuum limit to appear, we need to break away from the nome formalism and decouple $K$ from $K'$, effectively separating out the characteristic frequency $\Omega_0 \propto 1/K$, from the $K'$.

Defining $\Omega_0 = 2\pi/T_\text{prim}$, the above solutions partially collapse to:

\begin{equation}
	\begin{aligned}
		\text{Class 1: } & \; k < 1, \;\; q(\tau) = \frac{4 \Omega_0}{\sqrt{|A_0|}} \sum_{m \text{ odd}}^\infty \frac{q(k)^{m/2}}{1+q(k)^m}\cos[m \Omega_0 \tau] \\
		& \; k > 1, \;\; q(\tau) =  \frac{2 \Omega_0}{\sqrt{|A_0|}} \;\, \sum_{n=1}^\infty \frac{q(\frac{1}{k})^n}{1+q(\frac{1}{k})^{2n}}\cos[  n  \Omega_0 \tau]  + \frac{\Omega_0}{2\sqrt{|A_0|}} \\
		\text{Class 2: } & \; k < 1, \;\; q(\tau) = \frac{4 \Omega_0}{\sqrt{|A_0|}} \sum_{m \text{ odd}}^\infty \frac{q(k)^{m/2}}{1-q(k)^{m}} \sin[ m \Omega_0 \tau]
	\end{aligned}
\end{equation}

For the continuum limit to appear, the Fourier coefficients need to be written in the hyperbolic form to match exactly the Fourier sine/cosine transforms of the separatrix. Therefore,

\begin{equation}
	\begin{aligned}
		\text{Class 1: } & \; k < 1, \quad
		\frac{q(k)^{m/2}}{1+q(k)^m} & \longrightarrow \frac{1}{2 \cosh[\frac{m\pi K'(k)}{2 K(k)}]} & = \frac{1}{2 \cosh[\kappa \frac{m \Omega_0}{\Omega_L}]} \\
		& \; k > 1, \quad \frac{q(\frac{1}{k})^n}{1+q(\frac{1}{k})^{2n}} & \longrightarrow \frac{1}{2\cosh[\frac{n\pi K'(\frac{1}{k})}{K(\frac{1}{k})}]} & = \frac{1}{2 \cosh[\kappa \frac{n \Omega_0}{\Omega_L}]} \\
		\text{Class 2: } & \; k < 1, \quad \frac{q(k)^{m/2}}{1-q(k)^{m}} & \longrightarrow \frac{1}{2 \sinh[\frac{m\pi K'(k)}{2 K(k)}]} & = \frac{1}{2 \sinh[\kappa \frac{m \Omega_0}{\Omega_L}]}
	\end{aligned}
\end{equation}

Here, $\kappa = K'(k)$ is the complementary complete elliptic integral of the first kind. Note that $\kappa$ is a well-defined real value even when $k > 1$. Substituting the Fourier kernels back into the solutions yields:

\begin{equation}
	\begin{aligned}
		\text{Class 1: } & \; k < 1, \;\; q(\tau) =  \sum_{m \text{ odd}}^\infty \frac{2 \Omega_0}{\sqrt{|A_0|}\cosh(\kappa \frac{m \Omega_0}{\Omega_L})}\cos[m \Omega_0 \tau] \\
		& \; k > 1, \;\; q(\tau) =  \;  \sum_{n=1}^\infty  \frac{\Omega_0}{\sqrt{|A_0|}\cosh(\kappa \frac{n \Omega_0}{\Omega_L})} \cos[  n  \Omega_0 \tau]  + \frac{\Omega_0}{2\sqrt{|A_0|}} \\
		\text{Class 2: } & \; k < 1, \;\; q(\tau) = \sum_{m \text{ odd}}^\infty \frac{2\Omega_0}{\sqrt{|A_0|}\sinh(\kappa \frac{m \Omega_0}{\Omega_L})} \sin[ m \Omega_0 \tau]
	\end{aligned}
\end{equation}

At this point, the $k > 1$ regime of Class 1 does not respect parity because its harmonics span every positive integer. Its Fourier coefficients and primitive period are also different from the other two regimes by exactly a factor of $2$.

Following Ref.~\cite{Chachiyo2025upl}, the symmetrical spectral structure can be obtained if we define the period of all regimes as:

\begin{equation}
	T = \frac{4 \Re[K(k)]}{\Omega_L} \longrightarrow \Omega_0 = 2\pi/T
\end{equation}

In the solutions above, only the $k > 1$ regime of Class 1 needs to be adjusted because the fundamental frequency is now twice that of the primitive. By making a replacement $\Omega_0 \rightarrow 2 \Omega_0$ for this regime, the universal kernel and parity selection for regimes naturally emerge all at once:

\begin{equation}
	\begin{aligned}
		\text{Class 1: } & \; k < 1, \;\; q(\tau) =   \sum_{m \text{ odd}}^\infty \frac{2 \Omega_0}{\sqrt{|A_0|}\cosh(\kappa \frac{m \Omega_0}{\Omega_L})}\cos[m \Omega_0 \tau], \\
		& \; k > 1, \;\; q(\tau) =   \sum_{m \text{ even}}^\infty  \frac{2 \Omega_0}{\sqrt{|A_0|}\cosh(\kappa \frac{m \Omega_0}{\Omega_L})} \cos[  m \Omega_0 \tau] + \frac{\Omega_0}{\sqrt{|A_0|}}, \\
		\text{Class 2: } & \; k < 1, \;\; q(\tau) =  \sum_{m \text{ odd}}^\infty \frac{2\Omega_0}{\sqrt{|A_0|}\sinh(\kappa \frac{m \Omega_0}{\Omega_L})} \sin[ m \Omega_0 \tau].
	\end{aligned}\label{eq:proof_freq_sols}
\end{equation}

\subsubsection{Separatrix as Continuum Limit}

Having reformulated the kernel into hyperbolic form and separated out the characteristic frequency, we can now directly manipulate $\Omega_0$. If letting $\Omega_0 \to 0$ yields precisely the Fourier sine/cosine transform of the separatrix solution, it will have been evident that the separatrix is indeed the continuum limit from the nearby oscillatory regimes.

We first prepare the separatrix solution from the Jacobi framework when $k \to 1$ and perform the Fourier sine/cosine transform. From Eq.~(\ref{eq:jacobi_sols}) and the tabulated limiting properties~\cite[\href{http://dlmf.nist.gov/22.5.ii}{$\S$22.5(ii)}]{NIST:DLMF}: 

\begin{equation}
	\begin{aligned}
		\text{Class 1: } & \; k \to 1, \;\; q(\tau)  =   \frac{\Omega_L}{\sqrt{|A_0|}\cosh(\Omega_L \tau)} && = \int_0^\infty d\Omega\, \frac{1}{\sqrt{|A_0|}\cosh(\frac{\pi \Omega}{2 \Omega_L})} \cos(\Omega \tau), \\
		\text{Class 2: }  & \; k \to 1, \;\; q(\tau) =   \frac{\Omega_L}{\sqrt{|A_0|}} \tanh(\Omega_L \tau) && = \int_0^\infty d\Omega \, \frac{1}{\sqrt{|A_0|}\sinh(\frac{\pi \Omega}{2 \Omega_L})} \sin(\Omega \tau).
	\end{aligned}\label{eq:proof_sep_transform}
\end{equation}

We now turn to the oscillatory regimes and let $\Omega_0 \to 0$. This is a standard procedure for converting a Fourier series with the coefficient $c_n$ (or $s_n$) into Fourier sine/cosine transform. Therefore, from the coefficients in Eq.~(\ref{eq:proof_freq_sols}):

\begin{equation}
	\begin{aligned}
		\text{Class 1: } & \; \Omega_0 \to 0, \;\; q(\tau) = \frac{1}{2} \times \frac{1}{2 \pi} \int_0^\infty d\Omega \, T c_m \cos(\Omega \tau) = \int_0^\infty d\Omega \, \frac{\cos (\Omega \tau)}{\sqrt{|A_0|}\cosh(\kappa \frac{\Omega}{\Omega_L})}, \\
		\text{Class 2: } & \; \Omega_0 \to 0, \;\; q(\tau) = \frac{1}{2} \times \frac{1}{2 \pi} \int_0^\infty d\Omega \, T s_m \sin(\Omega \tau) = \int_0^\infty d\Omega \, \frac{\sin (\Omega \tau)}{\sqrt{|A_0|}\sinh(\kappa \frac{\Omega}{\Omega_L})}.
	\end{aligned}\label{eq:proof_sep_limit}
\end{equation}

The right-hand-side of Eq.~(\ref{eq:proof_sep_transform}) and Eq.~(\ref{eq:proof_sep_limit}) match precisely  ($\kappa = \pi/2$ as $k \to 1$). Therefore, it is shown rigorously that the separatrix is where the frequency combs of the nearby oscillatory regimes dissolve into continuum.

\subsection{Torque-Free Rigid-Body Rotation}\label{sec:method_spacecraft}

Here, we study the spectral anatomy of the Dzhanibekov effect. For moment of inertia $I_1 > I_2 > I_3$, we begin with Euler's equations for the angular velocity $(\omega_1, \omega_2, \omega_3)$ and rescale spatial units so that the governing equations take asymmetric forms:

\begin{equation}
	\begin{aligned}
		\frac{d y_1}{dz} = - y_2 y_3, \quad  \frac{d y_2}{dz} = + y_3 y_1, \quad \frac{d y_3}{dz} = - y_1 y_2,
	\end{aligned}\label{eq:asym_euler}
\end{equation}
where the scaling relations are:
\begin{equation}
	\begin{aligned}
		& y_1 =  +\sqrt{\frac{(I_1-I_3)(I_1-I_2)}{I_2 I_3}} \omega_1,  \\
		&  y_2 = -\sqrt{\frac{(I_2-I_3)(I_1-I_2)}{I_1 I_3}} \omega_2, \\
		&  y_3 = +\sqrt{\frac{(I_2-I_3)(I_1-I_3)}{I_1 I_2}} \omega_3, 
	\end{aligned}
\end{equation}

This allows us to rescale the derived properties back to physical space   $(\omega_1, \omega_2, \omega_3, t)$, valid for arbitrary principal moment of inertia $(I_1, I_2, I_3)$. In this asymmetric space, there are two constants of motion:

\begin{equation}
	C_1 = y_2^2 + y_1^2, \quad  C_3 = y_2^2 + y_3^2.
\end{equation}

We then map Eq.~(\ref{eq:asym_euler}) to the quartic potential in Eq.~(\ref{eq:quartic_physical}), yielding:

\begin{equation}
	\begin{aligned}
		& y_1 \text{ axis: } \quad  c_4 = +\frac{1}{2}, \, c_2 = \frac{C_3 - 2 C_1}{2},\quad && E = \frac{1}{2}C_1 (C_3 - C_1)  \\	
		& y_2 \text{ axis: } \quad c_4 = -\frac{1}{2},\,  c_2 = \frac{C_1 + C_3}{2}\quad  && E = \frac{1}{2} C_1 C_3   \\
		& y_3 \text{ axis: } \quad c_4 = +\frac{1}{2},\,  c_2 = \frac{C_1 - 2 C_3}{2}\quad  && E = \frac{1}{2} C_3 (C_1 - C_3)  \\
	\end{aligned}
\end{equation}

To allow a more straightforward derivation, we use the solutions in the physical space as elaborated in Appendix \ref{sec:physical_solution}.   We proceed to evaluate the squared-amplitudes $A_{1,2}$, $\Omega_L$, and $k$. Because of the term $\sqrt{\cdots}$ in evaluating $A_{1,2}$, the algebra branches into two scenarios: $C_1 > C_3$ and $C_3 > C_1$:

\begin{equation}
	\begin{aligned}
		&	C_1 > C_3
		\begin{cases}
			y_1 \text{ axis Class 1: } A_1 = C_1, A_2 = C_1 - C_3, & \Omega_L^2 = C_3, k^2 = C_1/C_3,  \\
			y_2 \text{ axis Class 2: } A_1 = C_3, A_2 = C_1, & \Omega_L^2 = C_1, k^2 = C_3/C_1, \\
			y_3 \text{ axis Class 1: } A_1 = C_3, A_2 = C_3 - C_1, & \Omega_L^2 = C_1, k^2 = C_3/C_1,
		\end{cases} \\
		&	C_3 > C_1
		\begin{cases}
			y_1 \text{ axis Class 1: } A_1 = C_1, A_2 = C_1 - C_3, & \Omega_L^2 = C_3, k^2 = C_1/C_3,  \\
			y_2 \text{ axis Class 2: } A_1 = C_1, A_2 = C_3, & \Omega_L^2 = C_3, k^2 = C_1/C_3, \\
			y_3 \text{ axis Class 1: } A_1 = C_3, A_2 = C_3 - C_1, & \Omega_L^2 = C_1, k^2 = C_3/C_1,
		\end{cases}
	\end{aligned} \label{eq:proof_space_class}
\end{equation}

From the expression for the period $T = 4 \Re[K(k)]/\Omega_L$, the analytic continuation $K(k) = K(1/k)/k$ is applied when $k > 1$. Surprisingly, 
all three axes converge to an identical expression for the periods within their respective scenarios as follows:

\begin{equation}
	\begin{aligned}
		& C_1 > C_3 \begin{cases}
			y_1 \text{ axis: } (k > 1), \quad  T = \frac{4 \Re[K(k)]}{\Omega_L} \longrightarrow T = \frac{4 K(\sqrt{C_3/C_1})}{\sqrt{C_1}}, \\
			y_2 \text{ axis: } (k < 1), \quad  T = \frac{4 \Re[K(k)]}{\Omega_L} \longrightarrow T = \frac{4 K(\sqrt{C_3/C_1})}{\sqrt{C_1}}, \\
			y_3 \text{ axis: } (k < 1), \quad  T = \frac{4 \Re[K(k)]}{\Omega_L} \longrightarrow T = \frac{4 K(\sqrt{C_3/C_1})}{\sqrt{C_1}},
		\end{cases} \\
		& C_3 > C_1 \begin{cases}
			y_1 \text{ axis: } (k < 1), \quad  T = \frac{4 \Re[K(k)]}{\Omega_L} \longrightarrow T = \frac{4 K(\sqrt{C_1/C_3})}{\sqrt{C_3}}, \\
			y_2 \text{ axis: } (k < 1), \quad  T = \frac{4 \Re[K(k)]}{\Omega_L} \longrightarrow T = \frac{4 K(\sqrt{C_1/C_3})}{\sqrt{C_3}}, \\
			y_3 \text{ axis: } (k > 1), \quad  T = \frac{4 \Re[K(k)]}{\Omega_L} \longrightarrow T = \frac{4 K(\sqrt{C_1/C_3})}{\sqrt{C_3}},
		\end{cases}
	\end{aligned} \label{eq:proof_space_modulus}
\end{equation}

The results from the two scenarios can be combined by defining $C_\text{min}$ and $C_\text{max}$ as the smaller and larger of the $\{C_1, C_3\}$. Hence, the unified expression for all possible configurations is:

\begin{equation}
	T = \frac{4 K(\sqrt{C_\text{min}/C_\text{max}})}{\sqrt{C_\text{max}}}.
\end{equation}

Transforming $C_1$ and $C_3$ to the $(\omega_1, \omega_2, \omega_3)$ coordinates, and rewriting them in terms of the energy and angular momentum directly, we arrive at Eq.~(\ref{eq:C1C3}).

\section{Discussion}\label{sec:discussion}

Here, we offer a case study that cuts deep into the heart of mathematical physics. Specifically, the symmetrical spectral structure may not only persist in multiple physics models (i.e., pendulum-like and quartic), but also traverse the architecture of time itself (see supplementary material $\S2.2$ for the detailed derivation).

Consider the simple harmonic archetype $(\sigma,\gamma)=(0,+1)$ in Fig.~\ref{fig:potential}. Because of the potential $V(q) = +q^2$, the particle oscillates with the fundamental frequency:

\begin{equation}
	\Omega_0 = \sqrt{2}.
\end{equation}

This leads to two solutions, their initial conditions, and total energy range:

\begin{equation}
	\begin{aligned}
		& q(\tau) = \sqrt{\epsilon}\cos(\Omega_0 \tau), &&  \quad (q_0, \dot{q}_0) = (\sqrt{\epsilon}, 0), &  \quad  \epsilon > 0, \\
		& q(\tau) = \sqrt{\epsilon}\sin(\Omega_0 \tau), &&  \quad (q_0, \dot{q}_0) = (0,\sqrt{2\epsilon}), &  \quad \epsilon > 0.
	\end{aligned}\label{eq:simple_harmonic_sols}
\end{equation}

A Wick rotation is a well-known mathematical technique whereby the time parameter in the governing equation is replaced by its imaginary counterpart, $\tau \to \pm i \tau$. This fundamentally alters the original governing equation and its solutions. We find that the Wick rotation precisely maps the bounded $(0,+1)$ archetype into the unbounded $(0,-1)$ archetype. The solutions strictly preserve their trigonometric/hyperbolic forms as follows.

In the $(0, +1)$ archetype, the energy equation is

\begin{equation}
\left(\frac{dq}{d\tau}\right)^2 = 2\epsilon - 2(+q^2).
\end{equation}

Applying the Wick rotation $\tau \to \pm i\tau$ means $d\tau \to \pm i\,d\tau$, and the governing equation becomes

\begin{equation}
	\left(\frac{dq}{d\tau}\right)^2 = 2\epsilon - 2(-q^2),
\end{equation}

where the energy also maps $\epsilon \to -\epsilon$ from the original to ensure the resulting physical trajectories $q(\tau)$ remain real. The resulting dynamics is therefore the $(0,-1)$ archetype. 

To find the solutions, we apply the joint substitution ($\tau \to \pm i\tau$ and $\epsilon \to -\epsilon$) directly to the two harmonic solutions in Eq.~(\ref{eq:simple_harmonic_sols}), while choosing the phase of the Wick-rotation (i.e., 90-degree clockwise or counterclockwise) that preserves the initial position and velocity of the solutions. This leads to:

\begin{equation}
	\begin{aligned}
		& q(\tau) = \sqrt{-\epsilon}\cosh(\Omega_0 \tau), &&  \quad (q_0, \dot{q}_0) = (\sqrt{-\epsilon}, 0), &  \quad  \epsilon < 0, \\
		& q(\tau) = \sqrt{+\epsilon}\sinh(\Omega_0 \tau), &&  \quad (q_0, \dot{q}_0) = (0,\sqrt{+2\epsilon}), &  \quad \epsilon > 0.
	\end{aligned}\label{eq:unbound_harmonic_sols}
\end{equation}

Physically, the former branch describes a particle that starts from rest and accelerates outward, while the latter starts with a kinetic kick that allows it to cross the potential peak.

In this case study, the Wick rotation ($\tau \to \pm i\tau$) is an orthogonal projection of the symmetrical spectral structure into a complementary time architecture. The bound space (real time) and the unbound space (imaginary time) are two sides of the same topological coin.

\subsection*{The Fundamental Clock: From Phase to Runaway}

The clock never breaks or disappears. Its temporal identity ($\Omega_0 = \sqrt{2}$)  simply rotates 90 degrees in the complex plane, and changes how it describes the dynamics. It transitions from a measure of cyclic endurance (how long a period lasts) to a measure of dynamic instability (how fast the system accelerates away from the origin).

\begin{itemize}
	\item \emph{In Real Time (Bound):} The clock is strictly real ($\Omega_0 = \sqrt{2}$). It measures phase accumulation. It counts the periodic, cyclic returns of a trapped trajectory.
	\item \emph{In Imaginary Time (Unbound):} Under Wick rotation, the clock becomes strictly imaginary ($\Omega_0 \to i\Omega_0$). An imaginary frequency does not count cycles; it measures the rate of exponential runaway.
\end{itemize}

\subsection*{Parity: The Topological Invariant}

The solutions exhibit the exact mirroring of $\cos \leftrightarrow \cosh$ and $\sin \leftrightarrow \sinh$. This suggests that parity selection is a deeper topological invariant than boundedness. Parity is insensitive to whether the motion is trapped or runaway; it depends only on the interaction with the origin.

\begin{itemize}
	\item \emph{Even Parity ($\cos$ and $\cosh$):} Both cosine and cosh share the initial condition $\dot{q}_0 = 0$. In real time, the cosine reaches its maximum amplitude and falls back. In imaginary time, the cosh particle ($\epsilon < 0$) lacks the energy to cross the potential peak, bounces off the inverted barrier, and repels outward. Even parity represents a \emph{bounce} at the origin.
	
	\item \emph{Odd Parity ($\sin$ and $\sinh$):} Both sine and sinh share the initial condition $q_0 = 0$. They represent a trajectory that transits the origin with unimpeded velocity. In real time, the sine swings freely through the bottom of the potential well. In imaginary time, the sinh particle possesses a positive kinetic kick ($\epsilon > 0$) that allows it to shoot directly over the inverted barrier. Odd parity represents a \emph{transit} at the origin.
\end{itemize}

\subsection*{Continuum: The Unification Limit}

The continuum limit can be interpreted more deeply than a purely Fourier statement. It is the absolute unification of dynamical regimes. In the pure $(0, +1)$ bound harmonic archetype, there is no separatrix because the potential walls rise to infinity. However, traversing into the $(0, -1)$ archetype via Wick rotation instead generates a separatrix at exactly $\epsilon = 0$.

\begin{itemize}
	\item \emph{In Real Time (Bound):} Regimes are strictly distinguished by their discrete spectral organizations (e.g., odd harmonics versus even harmonics). At the separatrix, the fundamental clock stops ($\Omega_0 \to 0$), and these mutually exclusive spectral combs dissolve, merging into a single, continuous spectral kernel. The rigid boundary between distinct dynamical states completely vanishes.
	
	\item \emph{In Imaginary Time (Unbound):} Regimes manifest as the topological difference between a bounce (even parity, governed by $\cosh$) and a transit (odd parity, governed by $\sinh$). At the exact separatrix ($\epsilon = 0$), the distinction between these hyperbolic parity states completely collapses. They fuse into a single, purely exponential runaway motion: $q(\tau) \propto \cosh(\Omega_0 \tau) + \sinh(\Omega_0 \tau) = e^{\Omega_0 \tau}$.
\end{itemize}

Taken together, the Wick rotation thought experiment hints that the three spectral pillars---Clock, Parity, and Continuum---are fundamentally agnostic to the choice of temporal construct. Real time yields discrete Fourier resonances; imaginary time yields continuous hyperbolic divergence. Their physical descriptions change, but their underlying mathematical identities transcend the architecture of time itself.

\section{Conclusion and Outlook}\label{sec:conclusion}

In this work, we have shown that conservative symmetric quartic systems possess a symmetrical spectral organization. Although their motions appear as distinct dynamical regimes in the time domain, their frequency-domain structure is governed by a common principle: the regimes share a fundamental clock, obey parity selection rules, and approach the separatrix through a discrete-to-continuum transition.

The conventional time-domain view emphasizes the differences among regimes. Each regime has its own trajectory pattern, Jacobi elliptic representation, and, in some cases, its own primitive period. This makes the dynamics appear fragmented: oscillations, rotations, trapped motions, crossing motions, and separatrices are treated as separate mathematical cases.

The spectral view reveals a different organization. Once the correct clock is chosen, the regimes are no longer disconnected. They become different samplings of the same spectral architecture. Some regimes occupy odd harmonics, while others occupy even harmonics and may carry a DC component. The separatrix is not an exceptional case outside this structure; it is the limit in which the spectral comb becomes dense and turns into the Fourier transform of the nonperiodic separatrix motion. This perspective changes the interpretation of regime distinction in nonlinear dynamics. Regimes that look qualitatively different in the time domain can be understood as different redistributions of spectral weight in the frequency domain. The trajectory changes, but the underlying clock, envelope, and selection rules persist.

The torque-free rigid-body application illustrates this principle in a familiar physical setting. The classical Dzhanibekov condition identifies where the intermediate-axis flip occurs, while the present taxonomy reveals how the spectrum reorganizes: the angular-velocity components share one clock, occupy different parity channels, exchange DC bias between stable axes, and approach a continuum spectrum at the separatrix. Thus, the framework is not merely a reformulation of known elliptic-function solutions, but a frequency-domain taxonomy that exposes hidden order across dynamical regimes.

As explored in our mathematical case study, this framework may reveal deeper insights, transcending the distinctions between trapped and runaway motion, or real and imaginary time. Together, the Clock, Parity, and the Continuum form a symmetrical spectral triad for 1D dynamics. The Clock defines temporal identity. Parity dictates regime organization. The Continuum is the limit at which all regimes merge into unity. Under this broader spectral lens, the rhythm, organization, and unification limit remain invariant, irrespective of the particle trajectories, potential archetypes, or temporal constructs.

Together with the pendulum-like dynamics in the earlier exposition~\cite{Chachiyo2025upl}, the persistent spectral traits also invite the possibility that the symmetrical spectral structure encompasses an entire class of major physics models. Despite drastically different mathematical setups---$U(\phi) = A\cos{\phi}$ for the pendulum, and $U(x) = A x^4 + B x^2$ for the quartic---their kernels share curiously parallel forms. These are $1/\cosh$ in the pendulum and generalize to include $1/\sinh$ in the quartic. If we suppose that the two models are connected, the connections should expand to cover other physics models as well. By mathematically classifying these connections, we may fundamentally change the complex landscape in which we view dynamical regimes into a simple and elegant spectral architecture---a symmetrical spectral structure in conservative 1D dynamics.

\section*{Conflict of Interest}
The author has no conflicts to disclose.

\section*{Declaration of generative AI and AI-assisted technologies in the manuscript preparation process}
During the preparation of this report, I used Gemini in order to polish English language. After using this tool, I reviewed and edited the content as needed and I take full responsibility for its content.

\begin{acknowledgments}
	The author thanks Dr. Hathaithip Chachiyo for insightful discussions on the topology of real and imaginary time which led to the Wick-rotation analysis in this work.
\end{acknowledgments}


\section*{Data Availability}

All data and code necessary to reproduce the results in this work are publicly available at \newline https://github.com/teepanis/symmetrical-spectral-structure.

\appendix

\section{Scaling Constants}\label{sec:scaling}

In the nontrivial cases where neither $c_4$ nor  $c_2$ is zero, the scaling $\xi, \lambda, \nu$ are unique. When either or both are zero, however, one or two of the scalings become arbitrary, in which case we set them numerically to unity (with their respective physical units).

\begin{equation}
\begin{aligned}
&	c_4 \ne 0, c_2 \ne 0: && \quad   \xi = \sqrt{2|c_4|}/\lambda, && \lambda = \sqrt{ \left| c_2 \right|}, &&  \nu = \xi^2/\lambda^2, \\
&	c_4 = 0, c_2 \ne 0: && \quad  \xi = 1, && \lambda = \sqrt{|c_2|}, &&  \nu = \xi^2/\lambda^2, \\
& c_4 \ne 0, c_2 = 0: && \quad \xi = \sqrt{2 |c_4|}/\lambda, && \lambda = 1, &&  \nu = \xi^2/\lambda^2, \\
& c_4 = 0, c_2 = 0: && \quad \xi = 1, && \lambda = 1, &&  \nu = \xi^2/\lambda^2.
\end{aligned}
 \label{eq:nontrivial_scaling}
\end{equation}

\section{Solutions in Physical Space}\label{sec:physical_solution}

For direct engineering applications, we convert the solution from the dimensionless quartic backbone back into physical space in Eq.~(\ref{eq:quartic_physical}). For broader applicability, the governing equation is expressed as a first integral of motion. In applications involving a physical mass, $U(x) = c_4 x^4 + c_2 x^2$ and $E$ represent the mass normalized potential and total energy, respectively (i.e., $E = \text{[Energy]}/ m$). To find solutions for the bound, nontrivial motions, one begins by evaluating the squared-amplitudes:

\begin{equation}
	A_0 = -2 c_4, \quad  A_{1,2} = \frac{-c_2 \pm \sqrt{c_2^2 + 4 c_4 E}}{2 c_4}. \label{eq:roots_physical}
\end{equation}

The spectral parameters can then be computed exactly as described in Eq.~(\ref{eq:spectral_scale}) and (\ref{eq:spectral_params}). Unlike in the all dimensionless quartic backbone, these parameters now have the physical dimension when applicable. For example, $T$ has the unit of $\text{[Time]}$, while $\Omega_0$ is now $\text{[Time]}^{-1}$. Interestingly, $\zeta = \sqrt{2|c_4|} $ is also a product of the spatial scaling $\xi$ and temporal scaling $\lambda$ in Appendix \ref{sec:scaling}, $\zeta = \xi \lambda$. Therefore, the name \textquote{space-time scaling} is literally accurate under this linear transformation.

Finally, simply substituting $q \to x$ and $\tau \to t$  into Eq.~(\ref{eq:spectral_osc}) yields the solution $x(t)$ in physical space. In addition, the solutions $x(t)$ are consistent with the initial conditions: $\{x_0,v_0\} = \{   \sqrt{A_1},0\}$ for Class 1, and $\{0,\sqrt{2 E}\}$ for Class 2. For other conditions, the time can be shifted $x(t) \to x(t-t_0)$ with an additional overall sign $x \to \pm x$ applied when needed.

\bibliographystyle{vancouver} 
\bibliography{spectral_taxonomy_v4a.bib}


\end{document}